\documentstyle[preprint,aps]{revtex}

\newcommand{\pder}[1]{{\partial \over \partial{#1}}}
\newcommand\pall{\slash\kern-2pt \slash}
\newcommand\Det{{\rm Det}}
\newcommand\half{{1 \over 2}}
\newcommand\Tr{{\rm Tr}}
\newcommand\Intg{\int d^2\xi\sqrt{g}}
\newcommand\Intl{\int d^2\xi\,l }
\newcommand\cvd{\nabla}
\def\sitarel#1#2{\mathrel{\mathop{\kern0pt #1}\limits_{#2}}}

\begin{document}
\draft

\title{ Weyl Anomalies of Strings in Temporal Gauge
\footnote{Published in Prog.~Theor.~Phys. 93 (1995) 455}
}

\author{ Teruhiko Kawano}

\address{Department of Physics, Kyoto University, Kyoto 606-01, Japan
\footnote{
Present address: Uji Research Center, Yukawa Institute for
Theoretical Physics,\\
\hspace*{3.3cm}Kyoto University, Uji 611, Japan\\
\hspace*{3.3cm}e-mail address: kawano@yisun1.yukawa.kyoto-u.ac.jp}
}

\preprint{\vbox{\hbox{KUNS-1316}
                \hbox{HE(TH)~95/01}
                \hbox{January, 1995}}}

\maketitle
\begin{abstract}
We consider two-dimensional quantum gravity coupled to matters in the
temporal gauge, using the Polyakov path integral.
We show that the integration over the metric can be explicitly
performed under some plausible assumptions.
We also discuss that the critical dimensions in string theory may not
be determined in the temporal gauge.

\end{abstract}

\section{Introduction}

For the past ten years, string theory has been intensively studied as a
candidate of the unified theory. In the development it has been
revealed that string theory has too many classical vacua. Although we
expect that only one vacuum is selected quantum-mechanically, we can
never find the true vacuum if only perturbative approaches are used.
Thus, the framework beyond perturbation theory is required, and
string field theory should be one of
the strong candidates.

Much effort has been devoted to searching for a satisfactory string field
theory\cite{LC,SGL,WITTEN,HIKKO,Npoly,ZW}.
However, it is proved to be very difficult to construct it, especially
for closed strings.
Although the light-cone gauge string field
theory\cite{LC} is consistently formulated, the lack of manifest Lorentz
covariance makes it difficult to get an insight into the underlying
structure of string theory.
Therefore a theory with the manifest
Lorentz covariance is desired. Zwiebach has proposed such
a theory\cite{ZW}. At present, more researches seem to
be required to get non-perturbative information from his theory.

Recently, a new formulation has been proposed\cite{SFH} as
a second-quantized string theory with $0\leq c\leq1$.
Let us briefly explain
the main features of the $c=0$ case, for simplicity.
The field operators of a string are the creation operator $\Psi(l)^{\dag}$
and the annihilation operator $\Psi(l)$, which creates and annihilates
a loop with length $l$, respectively. These operators
satisfy the commutation relation
\begin{eqnarray}
 [ \Psi(l), \Psi(l')^{\dag} ] = \delta(l-l').
\end{eqnarray}
The Hamiltonian of this theory is given by
\begin{eqnarray}
{\cal H} &=& \int dldl'\,(l+l')\Psi(l)^{\dag}\Psi(l')^{\dag}\Psi(l+l')
\\
          && +\,g \int dldl'\,l\,l'\Psi(l+l')^{\dag}\Psi(l)\Psi(l')
\\
          && + \int dl\, \rho(l)\Psi(l),
\end{eqnarray}
where $g$ is the string coupling constant, and $\rho(l)$ is the
amplitude for the process in which a loop with length $l$ vanishes.
The vacuum state $|0\rangle$ is defined as
\begin{eqnarray}
          \Psi(l)|0\rangle = 0.
\end{eqnarray}
Then, the amplitude for $n$ loops can be expressed as
\begin{eqnarray}
\lim_{D\rightarrow\infty}
\langle0|\exp[-D{\cal H}]\Psi(l_1)^{\dag}\Psi(l_2)^{\dag}
\cdots\Psi(l_n)^{\dag}|0\rangle,
\end{eqnarray}
where $D$ is interpreted as the geodesic distance from the incident
$n$ loops, which
was first introduced in the transfer-matrix formalism of $2D$ quantum
gravity based on the dynamical triangulation\cite{GEOD}.
These amplitudes are proved to satisfy the Schwinger-Dyson equations
\cite{SDeq} in the matrix model.
Therefore, this theory reproduces all
the known results in the $c=0$ matrix model.

Since one of the main difficulties in constructing string field
theory is to decompose each of all amplitudes into a set of propagators
and elementary interaction vertices, we expect that this
formulation gives an alternative direction toward a satisfactory
critical string field theory.

It was shown that the Hamiltonian for the $c=0$ case can be constructed
directly from the transfer-matrix formalism in the dynamical
triangulation\cite{WT}.
The alternative derivation\cite{JR} was also given by the stochastic
quantization of the matrix model. It was discussed there that the
geodesic distance $D$ can be interpreted as the fictitious time in the
stochastic quantization.

To derive this Hamiltonian from the continuum theory based on the
Polyakov path integral, the temporal gauge was proposed\cite{TEMP}
as a gauge-fixing condition. These authors have
almost reproduced the Hamiltonian for the $c=0$ case.

When we consider the critical string theory, the continuum approach
seems to be more tractable than the others.  So, introducing
matter fields into the system of pure gravity considered
in the temporal gauge,
we intend to search for such a string field Hamiltonian.
For that purpose, we need to estimate the integration over the degrees
of freedom of the gravity sector, especially the shift function.

It is the purpose of the paper to demonstrate that we can explicitly
perform the path integration over the metric, under some plausible
assumptions, for cylinder amplitudes with and without matters.
As the first step toward the new direction, it is interesting
to consider how the critical dimensions emerges in the temporal gauge.
In the Polyakov path integral, most of progress has been made in the
conformal gauge, where the meaning of the critical dimensions is
clear; the central charge of the matters for which the Weyl
anomalies coming both from gravity and matter sectors cancel out
each other. Thus in this case  we can ignore the dynamical
degrees of freedom of gravity, the Liouville modes. However,
in the temporal gauge, it is not clear what corresponds
to the Liouville modes.
Therefore, it is interesting to investigate the Weyl anomalies
in this gauge, which is one of the main subjects
in the present paper.

This paper is organized as follows:
After we review the temporal gauge\cite{TEMP}
in sect.\ref{sec:ADM}, we will first consider pure gravity in
sect.\ref{sec:gravity}. We refine the calculation in the paper
in a more systematic manner;
in particular, it will be shown that integration
over the shift function $k(t,x)$ can be made, which is needed in the
next section. In sect.\ref{sec:matter}, introducing matters into the
system considered in sect.\ref{sec:gravity}, we will explicitly
compute the cylinder amplitude with matters, a propagator
of closed string\cite{PROP}. In sect.\ref{sec:Discu},
we will give the discussion based on the calculations
in the preceding sections.

\section{Temporal gauge}
\label{sec:ADM}

In the ADM decomposition, a metric $g_{mn}$ on a two-dimensional
surface with the coordinates $\xi^m=(\xi^0,\xi^1)=(t,x)$ is
parametrized as
\begin{eqnarray}
 [ g_{mn}(\xi) ]   &=&
\left[\begin{array}{cc}
     N(\xi)^2 + h(\xi)\,k(\xi)^2     &     h(\xi)\,k(\xi)
\\
              h(\xi)\,k(\xi)         &         h(\xi)
\end{array}\right],
\end{eqnarray}
where $N(\xi)$ is the lapse function, $k(\xi)$ the shift function,
and $h(t,x)$ the metric on the time slice at $t$.

The temporal gauge\cite{TEMP} is defined as
\begin{eqnarray}
N(\xi) &=& 1,
\label{eq:lapse}\\
\partial_1 h(\xi) &=& 0.
\label{eq:temporal}
\end{eqnarray}
This gauge condition is consistent with the transfer-matrix formalism
initiated in \cite{GEOD}. In fact, the first condition
(\ref{eq:lapse}) allows us to
regard the geodesic distance from the boundary directly as the time
coordinate $t$. Furthermore, since in the dynamical triangulation,
all the links of
triangles are assumed to have equal length, the loop boundaries are
also meshed with equal length, and this fact justifies the second
condition (\ref{eq:temporal}).

Integrating Eq.(\ref{eq:temporal}) and setting $h=l(t)^2$, we thus
have the following parametrization of the metric in the temporal gauge:
\begin{eqnarray}
 [ \bar g_{mn}(\xi) ]   &=&
\left[\begin{array}{cc}
                           1 + l(t)^2 k(t,x)^2     &     l(t)^2 k(t,x)
\\
                               l(t)^2 k(t,x)         &     l(t)^2
\end{array}\right],
\end{eqnarray}
where $l(t)$ can be interpreted as the loop length on the time slice
at $t$.

In this gauge-fixing condition, there remains the following residual
gauge symmetry at each time $t$:
\begin{eqnarray}
t &\rightarrow& t'=t,
\\
x &\rightarrow& x'=x-\alpha(t),
\end{eqnarray}
under which the metric $\bar g_{mn}$ in the temporal gauge transforms as
\begin{eqnarray}
\delta_{res.} l(t)&=& 0,
\\
\delta_{res.} k(t,x)&=& \left(\pder{t}-\pder{x} k(t,x)\right)\alpha(t).
\label{eq:zerok}
\end{eqnarray}
The generator of this transformation is given by
$v^m_{res.}\partial_m=\alpha\pder{x}$.

\section{Pure Gravity}
\label{sec:gravity}

In this section, we consider pure gravity in the temporal gauge.
In particular, we make an explicit integration over the shift function
$k(\xi)$. The manipulation we develop here will enable us to examine
the propagator of a string in the next section.

Consider a worldsheet with the topology of cylinder.
We call its two boundaries $C$ and $C'$.
On these boundaries, we impose boundary conditions on loop
length as
\begin{eqnarray}
l(t=0)=l,\qquad l(t=D)=l',
\end{eqnarray}
where we use $t \in [0, D]$ with $D$ the geodesic distance
between $C$ and $C'$, as we explained in introduction.
The Polyakov path integral for this amplitude is given by
\begin{eqnarray}
Z(l',l;D)
&=&\int {{\cal D}_g g_{mn} \over \mbox{Vol(diff.)}}
\exp\left[-\mu_0\int d^2\xi\sqrt{g}\right]
\nonumber\\
&&\delta\left(\int_C\sqrt{g_{mn}d\xi^md\xi^n}-l\right)
\delta\left(\int_{C'}\sqrt{g_{mn}d\xi^md\xi^n}-l'\right)
\nonumber\\
&&\delta\left(\int N(g_{mn})dt-D\right).
\label{eq:Z}
\end{eqnarray}

Since the integrand has the reparametrization invariance:
\begin{eqnarray}
\xi^m &\rightarrow& \xi'^m = \xi^m - v^m(\xi),
\\
\delta g_{mn}(\xi) &=& \cvd_mv_n + \cvd_nv_m,
\end{eqnarray}
we should factor out this gauge degrees of freedom
and will impose the temporal gauge.
In this gauge,
$\delta\left(\int N(g_{mn})dt-D\right)$
actually means $t\in[0,D]$, and
\begin{eqnarray}
\delta\left(\int_{C}\sqrt{g_{mn}d\xi^md\xi^n}-l\right)
&\rightarrow&\
\delta\left(l(t=0)-l\right),
\\
\delta\left(\int_{C'}\sqrt{g_{mn}d\xi^md\xi^n}-l'\right)
&\rightarrow&\
\delta\left(l(t=D)-l'\right),
\\
\exp\left[-\mu_0\int d^2\xi\sqrt{g}\right]
&\rightarrow&\
\exp\left[-\mu_0\int dt l(t)\right],
\end{eqnarray}
with $x\in[0,1]$.

It is useful to introduce two orthonormal tangent vectors
$e_{\bot}$ and $e_{\pall}$:
\begin{eqnarray}
(e^{m}_{\bot})= (1, -k), \qquad (e^{m}_{\pall})= (0, l^{-1}),
\end{eqnarray}
which are, respectively, in the normal and tangential directions to
time slices, and satisfy the following relations:
\begin{eqnarray}
&&e^{m}_{\bot}\,e^{n}_{\bot}\,\bar g_{mn}
= e^{m}_{\pall}\,e^{n}_{\pall}\,\bar g_{mn}
= 1,
\\
&&e^{m}_{\bot}\,e^{n}_{\pall}\,\bar g_{mn} = 0,
\\
&&e^{m}_{\bot}\,e^{n}_{\bot}\,+e^{m}_{\pall}\,e^{n}_{\pall}\,=\,\bar g^{mn}.
\end{eqnarray}
The basis $\{ E^{\bot}, E^{\pall} \}$ of the dual cotangent vectors to
$\{e_{\bot}, e_{\pall}\}$ is then given by
\begin{eqnarray}
(E^{\bot}_m) = ( 1, 0 ), \qquad (E^{\pall}_m) = ( kl, l ).
\end{eqnarray}
For a vector $V^m$, we define
\begin{eqnarray}
V^{\bot}= E^{\bot}_m\,V^m, \qquad V^{\pall}= E^{\pall}_m\,V^m.
\end{eqnarray}
For a cotangent vector $V_m$, we also define
\begin{eqnarray}
V_{\bot}= e_{\bot}^m\,V_m, \qquad V_{\pall}= e_{\pall}^m\,V_m.
\end{eqnarray}
Thus the differential operators $\partial_{\bot}$, $\partial_{\pall}$
in the normal and tangential directions, respectively, are given by
\begin{eqnarray}
\partial_{\bot}&=&e_{\bot}^m\partial_m=\partial_0-k(t,x)\partial_1,
\nonumber\\
\partial_{\pall}&=&e_{\pall}^m\partial_m={1 \over l(t)}\partial_1.
\end{eqnarray}
Note that their conjugate operators are given by
\begin{eqnarray}
\partial^{\dag}_{\bot} = -\left(\partial_\bot-\omega\right),
\qquad
\partial^{\dag}_{\pall} = -\partial_{\pall},
\\
\omega = \pder{x}k(t,x) - {1 \over l(t)}\pder{t}l(t)\,,
\end{eqnarray}
since the conjugation $\dagger$ should be taken here under the
inner product
$\langle f_1 | f_2 \rangle$ on the space
of the functions on the surface:
\begin{eqnarray}
\langle f_1 | f_2 \rangle &=& \int d^2\xi\sqrt{\bar g}\,f_1(\xi)\,f_2(\xi)
\\
                      &=& \int d^2\xi\, l(t)\, f_1(\xi)\,f_2(\xi).
\end{eqnarray}

The measure ${\cal D}_{\bar g}g_{mn}$ is defined by the following
norm on the space of infinitesimal deformation $\delta g_{mn}$
of metric:
\begin{eqnarray}
||\,\delta g_{mn}\,||^2_{\bar g}
&=& \int d^2\xi\sqrt{\bar g}\,\bar g^{mk}\bar g^{nl}
                                        \,\delta g_{mn}\delta g_{kl}
\\
&=& \int dt {\delta l(t)^2 \over l(t)}
   + {1 \over 2}\int d^2\xi\, l(t)\,\left(l(t)\delta k(t,x)\right)^2
\nonumber\\
&& + \int d^2\xi\, l(t)\,\left[\left(\partial_{\bot}\delta v^{\bot}\right)^2
              + \left(\partial_{\pall}\delta v^{\pall}\right)^2\right],
\end{eqnarray}
where
\begin{eqnarray}
\delta v^{\bot}=\delta v_0 -k(t,x)\delta v_1,
\qquad
\delta v^{\pall}={1 \over l(t)}\delta v_1.
\end{eqnarray}

The norm on the space of tangent vectors on the worldsheet is
given by
\begin{eqnarray}
||\,\delta v^m\,||^2_{\bar g}
&=& \int d^2\xi\sqrt{\bar g}\ \bar g_{mn}\,\delta v^m\delta v^n
\nonumber\\
&=& \int d^2\xi l(t)\ \left[\left(\delta v^{\bot}\right)^2
                              + \left(\delta v^{\pall}\right)^2\right].
\end{eqnarray}
This defines the measure ${\cal D}_{\bar g} v^m$ for the
generators of the reparametrization transformation
connected to the identity.

Changing variables $\{\delta g_{mn}\}$ into physical variables and
gauge degrees of freedom, we obtain
\begin{eqnarray}
{\cal D}_{\bar g} g_{mn}
= \prod_t {dl(t) \over \sqrt{l(t)}}
   {\cal D}_lk\,{\cal D}_lv^{\bot}\,{\cal D}_l\tilde v^{\pall}\,
   \Det^\half\left[\partial^{\dag}_{\bot}\partial_\bot\right]\,
   \Det^{'\half}\left[\partial^{\dag}_{\pall}\partial_{\pall}\right],
\end{eqnarray}
where we denote by $\tilde v^{\pall}$ the non-zero modes of  $v^{\pall}$ for
$\partial_{\pall}$, and the determinant with a prime only includes
non-zero modes.
Recall that the measure ${\cal D}_lk$ is defined by
\begin{eqnarray}
||\,\delta k \,||^2_l = \half \int d^2\xi l\left(l\delta k\right)^2.
\label{eq:measurek}
\end{eqnarray}

Noting that the generator $v^m_{res.}$ of the residual symmetry
\begin{eqnarray}
v^{\bot}_{res.} &=& 0,
\\
v^{\pall}_{res.} &=& l(t)\alpha(t),
\label{eq:resv}
\end{eqnarray}
is the zero mode for the differential operator
$\partial_{\pall}$,
we can decompose Vol(diff.) as follows:
\begin{eqnarray}
\mbox{Vol(diff.)}
&=& \int {\cal D}_{\bar g}v^m.
\nonumber\\
&=& \int {\cal D}_{l}v^{\bot}\,{\cal D}_{l}\tilde v^{\pall}\,
         {\cal D}_{l}v^{\pall}_{res.}.
\end{eqnarray}

Thus, if we divide the measure ${\cal D}_{\bar g}g_{mn}$ by
Vol(diff.), the gauge degrees of freedom $\int {\cal D}_{l}v^{\bot}\,
{\cal D}_{l}\tilde v^{\pall}$ are eliminated and we obtain
\begin{eqnarray}
Z(l',l;D)
=\int \prod_t {dl(t) \over \sqrt{l(t)}}\,
   \left\{{1 \over \int{\cal D}_{l}v^{\pall}_{res.}}\right\}\,
   {\cal D}_lk\,&&
   \Det^\half\left[\partial^{\dag}_{\bot}\partial_\bot\right]\,
   \Det^{'\half}\left[\partial^{\dag}_{\pall}\partial_{\pall}\right]\,
   \exp\left[-\mu_0\int dt l(t)\right]
\nonumber\\
  && \cdot\delta\left(l(t=0)-l\right)
     \delta\left(l(t=D)-l'\right).
\label{eq:midZ}
\end{eqnarray}

In order to estimate the determinant
$\Det^\half\left[\partial^{\dag}_{\bot}
\partial_\bot\right]$, we begin with computing the determinant
$\Det\left[\triangle_{\bar g}\right]$
of the Laplacian
$\triangle_{\bar g}\left[\,l;k\right]$ defined as
\begin{eqnarray}
\triangle_{\bar g}
&=& -{1 \over \sqrt{\bar g}}\,
                \partial_m\sqrt{\bar g}\bar {g^{mn}}\partial_n\,,
\\
&=& \partial^{\dag}_{\bot}\partial_{\bot}
     +\partial^{\dag}_{\pall}\partial_{\pall}.
\end{eqnarray}

For a generic metric $g_{mn}$, we define $\Det[\triangle_g]$ by the
following heat kernel regularization which respects the
reparametrization invariance:
\begin{eqnarray}
\ln\Det\triangle_g
= - \int^{\infty}_\epsilon {d\tau \over \tau}
    \Tr\exp\left[-\tau\triangle_g\right].
\end{eqnarray}
If we perform an infinitesimal Weyl rescaling
$g_{mn}\rightarrow e^{2\delta\sigma}g_{mn}$,
the determinant $\Det\left[\triangle_g\right]$ changes by
\begin{eqnarray}
\delta\ln\Det\triangle_g =
-2\Intg\,\delta\sigma\left({1 \over 4\pi\epsilon} + {1 \over 12\pi}R[g]
                                                  + O(\epsilon)\right),
\label{eq:weyl}
\end{eqnarray}
where $R[g]$ is the scalar curvature defined as
\begin{eqnarray}
R[g]\,&=&-\half \,g^{mn}\,R^{l}_{mln}.
\\
R^{l}_{mnk}&=&\partial_k\Gamma^{l}_{mn}-\partial_n\Gamma^{l}_{mk}
            +\Gamma^{p}_{mn}\Gamma^{l}_{kp}
            -\Gamma^{p}_{mk}\Gamma^{l}_{np}\,.
\\
\Gamma^{p}_{mn}&=&\half\, g^{pq}
\left(\partial_mg_{nq}+\partial_ng_{mq}-\partial_qg_{mn}\right).
\end{eqnarray}
Furthermore, if one metric $g_{mn}$ is related to another one
$\hat g_{mn}$ by
$g_{mn}=e^{2\sigma}\hat g_{mn}$,
then the associated scalar curvatures have the following relation:
\begin{eqnarray}
R[g]&=&\triangle_g\sigma+R[\hat g]\,e^{-2\sigma}
\\
    &=&e^{-2\sigma}\left(\,\triangle_{\hat g}\sigma+R[\hat g]\,\right).
\label{eq:R}
\end{eqnarray}
Using this equation (\ref{eq:R}) and Eq.(\ref{eq:weyl}),
we thus obtain
\begin{eqnarray}
\Det\triangle_g = \Gamma[g]
\exp\left[-{1 \over 4\pi\epsilon}\Intg
          -{1 \over 12\pi}\Intg\, R[g]{1 \over \triangle_g}R[g]\right],
\label{eq:detg}
\end{eqnarray}
where the quantity $\Gamma[g]$ should be invariant under both
the reparametrization transformation and the Weyl rescaling.

Now that we have the expression (\ref{eq:detg}) for the determinant
$\Det\triangle_g$ with a generic metric $g_{mn}$, let us return to the
temporal gauge. Since the first term in the exponent
in Eq.(\ref{eq:detg}) can be eventually absorbed into the cosmological
term, what we need to investigate are the second term in the exponent
and the factor $\Gamma[\bar g]$.

As for the second term which we will denote by $A[l; k]$, since
the scalar curvature $R[g]$ in the temporal gauge is expressed as
\begin{eqnarray}
R[\bar g] &=& R[\,l ; k]
\\
          &=& \left(\partial_{\bot}-\omega\right)\omega
\\
          &=& -\partial^{\dag}_{\bot}\omega,
\label{eq:RT}
\end{eqnarray}
the term $A[l; k]$ turns out to be
\begin{eqnarray}
A[\,l;k] &=& {1 \over 12\pi}
              \int d^{2}\xi \sqrt{\bar g}
              \, R[\bar g]\,{1 \over \triangle_{\bar g}}\,R[\bar g]
\nonumber\\
         &=& {1 \over 12\pi}\Intl\, \partial^{\dag}_{\bot}\omega\,
             {1 \over \partial^{\dag}_{\bot}\partial_{\bot}
                      +\partial^{\dag}_{\pall}\partial_{\pall}}\,
                                    \partial^{\dag}_{\bot}\omega\,.
\end{eqnarray}

Let us then consider $\Gamma[\bar g]$. Here, we would like to know the
dependence of this quantity on the loop length $l(t)$ and
the shift function
$k(t,x)$. It is known that
metrics $g_{mn}$ have three kinds of deformations;
one under the reparametrization, one under the Weyl
rescaling and one associated with the change of
Teichm${\rm\ddot u}$ller parameters.
As we mentioned above, since
$\Gamma[g]$ should be invariant under both the reparametrization
and the Weyl rescaling, it can only depend on
the Teichm${\rm\ddot u}$ller parameter, which is one-dimensional in
our case. The deformation $\delta g^{T}_{mn}$ of the metric associated with
the Teichm${\rm\ddot u}$ller parameter should satisfy the following
equations:
\begin{eqnarray}
g^{mn}\delta g^{T}_{mn}=0,
\\
\cvd^{n}\delta g^{T}_{mn}=0.
\label{eq:quad}
\end{eqnarray}
In the temporal gauge, the corresponding equations are written in the
following form
for the deformations of $\delta l(t)$ and
$\delta k(t,x)$:
\begin{eqnarray}
\left(\partial_{\bot}-2\omega+{\dot l(t) \over l(t)}\right)\delta k(t,x)=0,
\label{eq:quadta}
\\
l(t)\partial_{\pall}\delta k(t,x)=
\left(\pder{t}-2\omega\right){\delta l(t) \over l(t)},
\label{eq:quadtb}
\end{eqnarray}
where we denote by a dot $\cdot$ the differentiation with respect to $t$.

Now we try to find a solution of these equations (\ref{eq:quadta}),
(\ref{eq:quadtb}). Rewriting Eq.(\ref{eq:quadtb}), we have
\begin{eqnarray}
\left(\pder{t}+2{\dot l(t) \over l(t)}\right)
{\delta l(t) \over l(t)}=
\pder{x}\left(\delta k(t,x) + 2{\delta l(t) \over l(t)}k(t,x)\right).
\label{eq:quadtc}
\end{eqnarray}
{}From the boundary conditions for $\delta k(t,x)$ and $k(t,x)$:
\begin{eqnarray}
\delta k(t,x=0)=\delta k(t,x=1),
\\
k(t,x=0)=k(t,x=1),
\end{eqnarray}
we find that the L.H.S. and the R.H.S. of Eq.(\ref{eq:quadtc}) should
be equal to zero, since the terms of the L.H.S.
depend only on $t$.
 Solving these equations, we obtain
\begin{eqnarray}
{\delta l(t) \over l(t)}
&=&\lambda\, l(t)^{-2},
\\
\delta k(t,x) &=& -2\lambda\, l(t)^{-2}k(t,x) + c(t),
\end{eqnarray}
where $\lambda$ is a constant, and $c(t)$ is an arbitrary function
depending only on $t$. However, substituting these into
Eq.(\ref{eq:quadta}), we can verify that there is only
trivial solution; $\lambda=0$, $c(t)=0$.

Thus, we may think that the deformations $\delta l(t)$ and
$\delta k(t,x)$ have nothing to do with the
Teichm${\rm\ddot u}$ller parameter;
This parameter can only be related to the geodesic distance $D$ and
the loop lengths $l,l'$ of the initial and final states.
It is thus plausible that we assume the independence of $\Gamma[\bar g]$
on the loop length $l(t)$ and the shift function $k(t,x)$,
and we will write
\begin{eqnarray}
\Gamma[\bar g] = \Gamma[l',l;D].
\end{eqnarray}

In summary, the result is
\begin{eqnarray}
\Det\triangle_{\bar g} &=&
\Gamma[l',l;D]
\exp\left[-{1 \over 4\pi\epsilon}\int dt\,l(t) - A[\,l,k]\right],
\\
&=& \Gamma[l',l;D]
\exp\left[-{1 \over 4\pi\epsilon}\int dt\,l(t)
          -{1 \over 12\pi}\Intl\, \partial^{\dag}_{\bot}\omega\,
             {1 \over \partial^{\dag}_{\bot}\partial_{\bot}
                      +\partial^{\dag}_{\pall}\partial_{\pall}}\,
                                    \partial^{\dag}_{\bot}\omega\,\right].
\label{eq:detdel}
\end{eqnarray}

Next, we investigate the determinant $\Det^\half\left[\partial^{\dag}_{\bot}
\partial_\bot\right]$. To do so, we use the following relation for the
Laplacian $\triangle_{\bar g}$ in the temporal gauge\cite{TEMP}:
\begin{eqnarray}
\triangle_{\bar g}\left[\, \beta^{-1}l;k\right]
= \partial^{\dag}_{\bot}\partial_{\bot}
     +\beta^2\partial^{\dag}_{\pall}\partial_{\pall},
\end{eqnarray}
and thus define the determinant $\Det\left[\partial^{\dag}_{\bot}
\partial_\bot\right]$ by
\begin{eqnarray}
\ln\Det\left[\partial^{\dag}_{\bot}\partial_\bot\right]
= \lim_{\beta \rightarrow 0}
  \ln\Det\triangle_{\bar g}\left[\,\beta^{-1}l;k\right].
\end{eqnarray}

Here we will make use of Eq.(\ref{eq:detdel}) to estimate the above.
Since we can see from the expression (\ref{eq:RT}) that
the scalar curvature
$R[\bar g]$ is invariant under the constant rescaling of loop
length $l(t)$:
\begin{eqnarray}
R[\,\beta^{-1}l; k] = R[\,l; k]
\qquad ( \beta~ \mbox{is a constant}),
\end{eqnarray}
we obtain
\begin{eqnarray}
\lim_{\beta \rightarrow 0}A[\,\beta^{-1}l;k]
&=& \lim_{\beta \rightarrow 0}{1 \over 12\pi\beta}\Intl\,
   \partial^{\dag}_{\bot}\omega\,
    {1 \over \partial^{\dag}_{\bot}\partial_{\bot}
      +\beta^2\partial^{\dag}_{\pall}\partial_{\pall}}\,
                                    \partial^{\dag}_{\bot}\omega\
\nonumber\\
&=& \lim_{\beta \rightarrow 0}{1 \over 12\pi\beta}\Intl\, \omega^2
\nonumber\\
&=& \lim_{\beta \rightarrow 0}{1 \over 12\pi\beta}
   \left[\,\int dt {{\dot l(t)}^2 \over l(t)}
    + \Intl\,\left(l\partial_{\pall}k\right)^2\,\right].
\label{eq:beta0}
\end{eqnarray}

As for the factor $\Gamma[l',l;D]$,
we first suppose to investigate the determinant
$\Delta_{\bar g}$ with
loop lengths $\beta l$ and $\beta l'$
at initial and final time,
respectively.
After that, we scale loop length $l(t)$ as
$l(t)\rightarrow\beta^{-1}l(t)$. To this end, the factor
$\Gamma[l',l;D]$ has the loop lengths $l,l'$ at the initial
and final time, respectively.

Thus we obtain
\begin{eqnarray}
\Det&&\left[\partial^{\dag}_{\bot}\partial_\bot\right]
= \lim_{\beta \rightarrow 0}
\Gamma[l',l;D]
\exp\left[-{1 \over 4\pi\beta\epsilon}\int dt\,l(t)
                                - A[\,\beta^{-1}l,k]\right],
\nonumber\\
&&= \lim_{\beta \rightarrow 0}
\Gamma[l',l;D]
\exp\left[-{1 \over 4\pi\beta\epsilon}\int dt\,l(t)
          - {1 \over 12\pi\beta}\left\{\int dt {{\dot l(t)}^2 \over l(t)}
          + \Intl\,\left(l\partial_{\pall}k\right)^2\,\right\}\right].
\label{eq:det}
\end{eqnarray}

We substitute Eq.(\ref{eq:det}) into Eq.(\ref{eq:midZ}).
Renormalizing the first term of exponent in Eq.(\ref{eq:det}) into
the bare cosmological constant $\mu_0$, we denote the renormalized
cosmological constant by $\mu$. Then, we have
\begin{eqnarray}
Z(l',l;D)
=
\Gamma[l',l;D]^{\half}
\lim_{\beta \rightarrow 0}
\int \prod_t &&{dl(t) \over \sqrt{l(t)}}\,
   \left\{{1 \over \int{\cal D}_{l}v^{\pall}_{res.}}\right\}\,
   {\cal D}_lk\,
    \Det^{'\half}\left[\partial^{\dag}_{\pall}\partial_{\pall}\right]\,
    \exp\left[-\mu\,\int dt l(t)\right]
\nonumber\\
&& \cdot
\exp\left[- {1 \over 24\pi\beta}\left\{\int dt {{\dot l(t)}^2 \over l(t)}
          + \Intl\,\left(l\partial_{\pall}k\right)^2\,\right\}\right]
\nonumber\\
  && \cdot\delta\left(l(t=0)-l\right)
     \delta\left(l(t=D)-l'\right).
\label{eq:lateZ}
\end{eqnarray}

As we mentioned in the last section, we have to further fix the residual
symmetry. We can see this from the fact that the zero mode $\delta
k_0(t)$ satisfying $\partial_{\pall}\delta k(t,x)=0$ does not appear
in the integrand of the R.H.S. of Eq.(\ref{eq:lateZ}).
{}From Eq.(\ref{eq:zerok}) and Eq.(\ref{eq:resv}), the following
relation is obtained:
\begin{eqnarray}
\delta_{res.} k(t,x)
= \left(\pder{t}-k'(t,x)\right)
{1 \over l(t)}\delta v^{\pall}_{res.}(t),
\end{eqnarray}
where we denote by a prime $\prime$ the differentiation with respect
to $x$.
Therefore, we apply the Fadeev-Popov prescription to it;
namely, we substitute the identity
\begin{eqnarray}
1=\int{\cal D}_l v^{\pall}_{res.}\prod_{t}l(t)^{-\half}
&&\prod_{t}\delta\left(k(t,x_0)-\left(\pder{t}-k'(t,x_0)\right)
{1 \over l(t)}\delta v^{\pall}_{res.}(t)\right)
\nonumber\\
&&\cdot
\Det^{-1}\left[\left(\pder{t}-k'(t,x_0)\right){1 \over l(t)}\right]
\end{eqnarray}
into Eq.(\ref{eq:lateZ}), where $x_0$ is an arbitrary value of $x$.
Furthermore we decompose the measure ${\cal D}_lk$ into the part of the
zero mode ${\cal D}_lk_0$ and the part of the non-zero mode
${\cal D}_l\tilde k$. From the definition (\ref{eq:measurek}) of the
measure ${\cal D}_lk$, we can verify that
\begin{eqnarray}
\int{\cal D}_lk_0 \cdot\,1
&=&\int\prod_{t}dk_0(t)\prod_{t}l(t)^{{3 \over 2}} \cdot\,1
\\
&=&\int{\cal D}_lv^{\pall}_{res.}\,\prod_{t}l(t)^2\,
\Det^{-1}\left[\left(\pder{t}-k'(t,x_0)\right)\right].
\label{eq:FP}
\end{eqnarray}
Accordingly, the volume factor
$\int{\cal D}_lv^{\pall}_{res.}$ of the residual symmetry
in Eq.(\ref{eq:lateZ}) and the volume factor
$\int{\cal D}_lv^{\pall}_{res.}$ emerging from
Eq.(\ref{eq:FP}) cancel out.
Thus,
\begin{eqnarray}
Z(l',l;D)
=\lim_{\beta \rightarrow 0}
\Gamma[l',l;D]^{\half}
\int \prod_t &&dl(t)l(t)^{3 \over 2}\,
     {\cal D}_l\tilde k\,
\Det^{-1}\left[\left(\pder{t}-k'(t,x_0)\right)\right]
\nonumber\\
&&\cdot
\exp\left[- {1 \over 24\pi\beta}\left\{\int dt {{\dot l(t)}^2 \over l(t)}
          + \Intl\,\left(l\partial_{\pall}k\right)^2\,\right\}\right]
\nonumber\\
  && \cdot\Det^{'\half}\left[\partial^{\dag}_{\pall}\partial_{\pall}\right]\,
   \exp\left[-\mu\,\int dt l(t)\right]
\nonumber\\
  && \cdot\delta\left(l(t=0)-l\right)
     \delta\left(l(t=D)-l'\right).
\label{eq:latterZ}
\end{eqnarray}

Since the exponent of the R.H.S. in Eq.(\ref{eq:latterZ}) means that
$\partial_{\pall}k=0$ in the limit $\beta\rightarrow 0$, we can
ignore $k'$ in the determinant $\Det^{-1}[(\pder{t}-k'(t,x_0))]$ in
the same equation. It is thus easy to perform the integration over
the non-zero mode $\tilde k$. This yields
$\Det^{'-\half}[{1 \over 12\pi\beta}\partial^{\dag}_{\pall}
\partial_{\pall}]$, which cancels
$\Det^{'\half}[\partial^{\dag}_{\pall}\partial_{\pall}]$
in the R.H.S. of (\ref{eq:latterZ}).

After all, we obtain
\begin{eqnarray}
Z(l',l;D)
=\lim_{\beta \rightarrow 0}\tilde\Gamma[l',l;D;\beta]^{\half}
\int \prod_t l(t)^{3 \over 2}\,dl(t)\,&&
\exp\left[-\mu\,\int dt l(t)
          - {1 \over 24\pi\beta}\int dt {{\dot l(t)}^2 \over l(t)}\right]
\nonumber\\
  && \cdot\delta\left(l(t=0)-l\right)
     \delta\left(l(t=D)-l'\right),
\label{eq:finalZ}
\end{eqnarray}
where $\tilde\Gamma[l',l;D;\beta]^{\half}
=\Gamma[l',l;D]^{\half}\Det^{-1}[\pder{t}]\Det^{\half}[12\pi\beta]$.

Note that the power of the loop $l(t)$, apart from those
exponentiated, in Eq.(\ref{eq:finalZ}) is different from that in
\cite{TEMP}.
Ours is three half, while theirs is minus one.
This discrepancy will be discussed in sect.\ref{sec:Discu}.

\section{Propagator}
\label{sec:matter}

We now introduce scalar fields into the system considered in the
last section. In particular, we pay attention to what corresponds to
the Weyl anomalies in this case, which appear in the conformal gauge.

We substitute the path integral for $N$ scalar fields (string
coordinates) $X^{\mu}(\xi)$ $(\mu=1,\cdots,N)$
\begin{eqnarray}
W[g] = \int{\cal D}_gX e^{-S[X,g]}
\label{eq:W}
\end{eqnarray}
into the amplitude (\ref{eq:Z}) in the last section.
Here the action $S[X,g]$ is given by
\begin{eqnarray}
S[X,g]={1 \over 8\pi}\int_{\Sigma}d^2\xi
         \sqrt{g}g^{mn}\partial_mX^{\mu}\partial_nX^{\mu}.
\end{eqnarray}
This action describes a string propagating
in the $N$ dimensional flat Euclidean space-time. So the amplitude under
consideration can be regarded as a propagator of
such a string\cite{PROP}.

We have to impose boundary conditions on the scalar fields
$X^{\mu}(\xi)$ at the boundaries $C$ and $C'$.
Since the string coordinates $X^{\mu}(\xi)$ map the worldsheet into
the space-time, if two string coordinates can be connected under
the reparametrization transformation on the worldsheet, we should
regard these as the same string configuration.
Thus, up to the reparametrizations, we specify the
boundary conditions as follows:
\begin{eqnarray}
X^{\mu}(t=0,x)&=&X^{\mu}_i(x) \qquad (\, \mbox{on}~ C\,  ),
\label{eq:bcMi}\\
X^{\mu}(t=D,x)&=&X^{\mu}_f(x) \qquad (\, \mbox{on}~ C'\, ).
\label{eq:bcMf}
\end{eqnarray}

Then, the string propagator $G(l',X_f\,;l,X_i\,;D)$ is given by
\begin{eqnarray}
G(l',X_f\,;l,X_i\,;D)
&=&\int d\Sigma^{\mbox{diff.}}_{i,f}
\int {{\cal D}_g g_{mn} \over \mbox{Vol(diff.)}}
\exp\left[-\mu_0\int d^2\xi\sqrt{g}\right]W[g;X_f,X_i]
\nonumber\\
&&\cdot\delta\left(\int_C\sqrt{g_{mn}d\xi^md\xi^n}-l\right)
\delta\left(\int_{C'}\sqrt{g_{mn}d\xi^md\xi^n}-l'\right)
\nonumber\\
&&\cdot\delta\left(\int N(g_{mn})dt-D\right),
\label{eq:G}
\end{eqnarray}
where $d\Sigma^{\mbox{diff.}}_{i,f}$ denotes integration over the
reparametrizations on the boundaries $C$, $C'$.
Furthermore, for the path integral $W[g]$
over the string coordinates, we
explicitly represent its dependence on the boundary conditions of
$X^{\mu}(\xi)$ as $W[g;X_f,X_i]$.

Let us first compute $W[g;X_f,X_i]$.
Let $\bar X^{\mu}_g$ be the solution of the equation of motion
\begin{eqnarray}
\Delta_g\bar X^{\mu}_g = 0
\label{eq:path}
\end{eqnarray}
satisfying the above boundary conditions (\ref{eq:bcMi},\ref{eq:bcMf}).
Then we expand the string coordinates $X^{\mu}(\xi)$
around the solution $\bar X^{\mu}_g(\xi)$ as
\begin{eqnarray}
X^{\mu}(\xi)=\bar X^{\mu}_g(\xi) +y^{\mu}(\xi),
\label{eq:expanX}
\end{eqnarray}
and substitute these into the path integral $W[g;X_f,X_i]$.
Integration now are made
over the variables $y^{\mu}(\xi)$
satisfying the boundary conditions:
\begin{eqnarray}
y^{\mu}(t=0,x)=y^{\mu}(t=D,x)=0,
\end{eqnarray}
and the measure is defined as
\begin{eqnarray}
||\delta y^{\mu}||^2_g
= {1 \over 8\pi}\Intg\, \delta y^{\mu}(\xi)\,\delta y^{\mu}(\xi).
\end{eqnarray}
Furthermore, the action $S[X,g]$ turns out to be
\begin{eqnarray}
S[X,g]= S_{{cl.}} + S[y,g],
\end{eqnarray}
where the classical action $S_{{cl.}}$ is
\begin{eqnarray}
S_{{cl.}} &=& S[\bar X^{\mu}_g,g]
\nonumber\\
&=& {1 \over 8\pi}\int dx \left[\bar X^{\mu}_g(\xi)\,
\sqrt{g}\,g^{0n}\,\partial_n\bar X^{\mu}_g(\xi)\right]^{t=D}_{t=0}.
\end{eqnarray}
Thus, the path integral $W[g;X_f,X_i]$ is easily performed, and we obtain
\begin{eqnarray}
W[g;X_f,X_i]&=& e^{-S_{{cl.}}}\int{\cal D}_gy
\exp\left[-{1 \over 8\pi}\Intg y^{\mu}(\xi)\Delta_gy^{\mu}(\xi)\right]
\\
&=&e^{-S_{{cl.}}}\left(\Det\Delta_g\right)^{-{N \over 2}}.
\end{eqnarray}

Imposing the temporal gauge on the path integral $W[g;X_f,X_i]$
and using Eq.(\ref{eq:detdel}) in the last section,
we find
\begin{eqnarray}
&&W[\bar g;X_f,X_i]
\nonumber\\
&&\ = \Gamma[l',l;D]^{-{N \over 2}}\,e^{-S_{{cl.}}}\,
\exp\left[-{N \over 8\pi\epsilon}\int dt\,l(t)
          -{N \over 24\pi}\Intl\, \partial^{\dag}_{\bot}\omega\,
             {1 \over \partial^{\dag}_{\bot}\partial_{\bot}
                      +\partial^{\dag}_{\pall}\partial_{\pall}}\,
                                    \partial^{\dag}_{\bot}\omega\,\right].
\label{eq:Wt}
\end{eqnarray}
The second term in the exponent corresponds to the Weyl anomaly from
the matters in the conformal gauge.

Substituting this equation (\ref{eq:Wt}) into the string propagator
(\ref{eq:G}) and rewriting the remaining part
in a similar way as we did in the last section,
we can verify that
\begin{eqnarray}
G(l',X_f\,;l,X_i\,;D)
=\lim_{\beta \rightarrow 0}&&\int d\Sigma^{\mbox{diff.}}_{i,f}
\int \prod_t dl(t)l(t)^{3 \over 2}\,
     {\cal D}_l\tilde k\,
    \Gamma[l',l;D]^{{1-N \over 2}}\,e^{-S_{{cl.}}}\,
\nonumber\\
&&\cdot
\exp\left[- {1 \over 24\pi\beta}\left\{\int dt {{\dot l(t)}^2 \over l(t)}
          + \Intl\,\left(l\partial_{\pall}k\right)^2\,\right\}\right]
\nonumber\\
&&\cdot
\exp\left[-\mu_N\int dt\,l(t)
          -{N \over 24\pi}\Intl\, \partial^{\dag}_{\bot}\omega\,
             {1 \over \partial^{\dag}_{\bot}\partial_{\bot}
                      +\partial^{\dag}_{\pall}\partial_{\pall}}\,
                                    \partial^{\dag}_{\bot}\omega\,\right]
\nonumber\\
&&\cdot
\Det^{-1}\left[\left(\pder{t}-k'(t,x_0)\right)\right]
\Det^{'\half}\left[\partial^{\dag}_{\pall}\partial_{\pall}\right]\,
\nonumber\\
  && \cdot\delta\left(l(t=0)-l\right)
     \delta\left(l(t=D)-l'\right),
\label{eq:P}
\end{eqnarray}
where we denote by $\mu_N$ the renormalized cosmological constant.
In this expression, the classical action $S_{cl.}$ is
\begin{eqnarray}
S_{cl.}
= {1 \over 16\pi}\int dx \left[ \pder{t}\left\{l(t)
\left(\bar X^{\mu}_g\right)^2\right\}
+\omega l(t)\left(\bar X^{\mu}_g\right)^2\right]^{t=D}_{t=0}.
\end{eqnarray}

The terms multiplied by $\beta^{-1}$ in the exponent in the R.H.S.
of Eq.(\ref{eq:P}) mean that the following configuration dominates
in the limit $\beta\rightarrow 0$:
\begin{eqnarray}
\dot l(t) &=& 0,
\\
k'(t,x) &=& 0.
\end{eqnarray}
Therefore,
\begin{eqnarray}
\omega &\sitarel{\longrightarrow}{\beta\rightarrow 0}& 0,
\\
\Delta_g &\sitarel{\longrightarrow}{\beta\rightarrow 0}&
                     -\left(\partial_0-k_0(t)\partial_1\right)
                       \left(\partial_0-k_0(t)\partial_1\right)
                     -l(t)^{-2}\partial_1\partial_1,
\end{eqnarray}
where $k_0(t)$ is the zero mode of $k(t,x)$ for the differential
operator $\partial_{\pall}$. As we can see from these equations,
the above-mentioned Weyl anomalies from the scalar fields vanish in
this limit; namely
\begin{eqnarray}
-{N \over 12\pi}\Intl\, \partial^{\dag}_{\bot}\omega\,
             {1 \over \partial^{\dag}_{\bot}\partial_{\bot}
                      +\partial^{\dag}_{\pall}\partial_{\pall}}\,
                                    \partial^{\dag}_{\bot}\omega\
\sitarel{\longrightarrow}{\beta \rightarrow 0} 0.
\end{eqnarray}

By similarly calculating the remaining part in the string propagator
(\ref{eq:P}) and integrating out the non-zero mode $\tilde k(t,x)$,
the following result is obtained:
\begin{eqnarray}
G(l',X_f\,;l,X_i\,;D)
=\lim_{\beta \rightarrow 0}&&\int d\Sigma^{\mbox{diff.}}_{i,f}
\int \prod_t dl(t)l(t)^{3 \over 2}\,
\tilde \Gamma[l',l;D]^{{1-N \over 2}}\,e^{-S_{{cl.}}}\,
\nonumber\\
&&\cdot
\exp\left[-\mu_N\int dt\,l(t)
- {1 \over 24\pi\beta}\int dt {{\dot l(t)}^2 \over l(t)}\right]
\nonumber\\
  && \cdot\delta\left(l(t=0)-l\right)
     \delta\left(l(t=D)-l'\right).
\label{eq:finalP}
\end{eqnarray}
where $\tilde \Gamma[l',l;D]^{{1-N \over 2}}=
\Gamma[l',l;D]^{{1-N \over 2}}\Det^{\half}[12\pi\beta]
\Det^{-1}[\pder{t}]$, and the classical action $S_{cl.}$ is
\begin{eqnarray}
S_{cl.}
= {1 \over 16\pi}\int dx \left[ l(t) \pder{t}
\left(\bar X^{\mu}_g\right)^2\right]^{t=D}_{t=0}.
\end{eqnarray}

\section{Discussion}
\label{sec:Discu}

In this paper we have considered two-dimensional quantum gravity in
the temporal gauge and
have demonstrated that we can explicitly perform the path integration
over the metric under some plausible assumptions.

In sect.\ref{sec:gravity}, we investigated the cylinder amplitude for
pure gravity in a different way from that in \cite{TEMP}.
As we mentioned at the end of that section, the discrepancy
between their result and ours (\ref{eq:finalZ}) was found in
the power of the loop length $l(t)$,
apart from those exponentiated.
This discrepancy may be explained as a difference in the
way to fix the residual symmetry.

Eq.(\ref{eq:finalZ}) implies that the cylinder amplitude
is essentially proportional to the delta function $\delta(l-l')$ in the
limit $\beta\rightarrow 0$. Then the loop length $l(t)$ should be
replaced by the one $l$ at the initial state.
So it is not clear how relevant this discrepancy is, until we can
compute the function $\Gamma[l',l; D]$.

In sect.\ref{sec:matter}, we considered a propagator of a string
propagating on the $N$ dimensional flat Euclidean space-time.
There we have been able to derive what should correspond to
the Weyl anomalies from
the matters, which finally vanishes in the limit $\beta\rightarrow 0$.
There are two subtleties in this calculation: first, it is not clear
how to integrate over the reparametrizations on the boundaries $C$ and
$C'$. Secondly, the validity of our assumption made on
$\Gamma[l,l';D]$ should be examined. As for the latter,
it is necessary to establish
how the Teichm${\rm\ddot u}$ller parameter depends on the geodesic
distance $D$ and the loop length $l,l'$ of the initial and final
states.

Despite these subtleties in this approach,
there seems no critical dimensions in the temporal gauge,
as we can see from Eq.(\ref{eq:finalP}).
However, to reach a decisive conclusion as to whether there really
exists no critical dimension in the temporal gauge approach, further
investigation is needed on the above-mentioned problems. Furthermore,
if it turns out to be the case, it is very interesting to examine the mass
spectrum of the physical states, especially the graviton ones.

\begin{acknowledgements}

The author would like to thank
M.~Fukuma, K.~Itoh, T.~Kugo and M.~Ninomiya
for valuable discussions and comments.
He is especially grateful to
M.~Fukuma and M.~Ninomiya
for their critical reading of the manuscript.

\end{acknowledgements}

\end{document}